\documentclass[12pt]{article}
\input epsf.sty
\usepackage{epsfig}
\newcommand{\beq}{\begin{eqnarray}}
\newcommand{\eeq}{\end{eqnarray}}

\newcommand{\be}{\begin{equation}}
\newcommand{\ee}{\end{equation}}
\newcommand{\ben}{\begin{eqnarray}\displaystyle}
\newcommand{\een}{\end{eqnarray}}

\def\sqr#1#2{{\vcenter{\vbox{\hrule height.#2pt
         \hbox{\vrule width.#2pt height#1pt \kern#1pt
            \vrule width.#2pt}
         \hrule height.#2pt}}}}

\begin{document}

{}~ \hfill\vbox{\hbox{hep-ph/0408251} \hbox{PUPT-2133} }\break

\vskip 1cm

\begin{center}
\Large{\bf Do Chiral Soliton Models Predict Pentaquarks?}\footnote{Contribution to the proceedings of the QCD 2004 Workshop at the University of Minnesota, May 13-16, 2004.}

\vspace{20mm}

\normalsize{Igor R. Klebanov and Peter Ouyang}

\vspace{10mm}

\normalsize{\em Joseph Henry Laboratories, Princeton University,}

\vspace{0.2cm}

\normalsize{\em Princeton, New Jersey 08544, USA}
\end{center}

\vspace{10mm}

\begin{abstract}

\medskip

We reconsider the relationship between the bound state and the $SU(3)$ rigid rotator approaches to strangeness in chiral soliton models. For non-exotic $S=-1$ baryons the bound state approach matches for small $m_K$ onto the rigid rotator approach, and the bound state mode turns
into the rotator zero-mode. However, for small $m_K$, there are no
$S=+1$ kaon bound states or resonances in the spectrum. 
This shows that for large $N$ and small
$m_K$ the exotic state is an artifact of the rigid rotator
approach. An $S=+1$ near-threshold state with the quantum numbers of
the $\Theta^+$ pentaquark comes
into existence only when sufficiently strong
$SU(3)$ breaking is introduced into the chiral lagrangian. 
Therefore, pentaquarks are not generic predictions of the chiral soliton
models.
\end{abstract}

\section{Introduction}

These lecture notes are largely based on our paper with N. Itzhaki and L. Rastelli\cite{us}.

Recently there has been a flurry of research activity on exotic
pentaquark baryons, prompted by reports\cite{Nakano,Stepanyan,Barth}
of the observation of the $S=+1$ baryon $\Theta^+$ (1540).  The
original photoproduction experiment\cite{Nakano} was largely motivated
by theoretical work\cite{DPP} in which chiral soliton models were used
to predict a rather narrow $I=0$, $J^P=\frac{1}{2}^+$ exotic $S=+1$
baryon whose minimal quark content is $uudd\bar{s}$. The method used
in \cite{DPP} to predict the baryon spectrum is the $SU(3)$ collective
coordinate quantization of chiral solitons\cite{Witten,Guad}.  This
approach predicts the well-known ${\bf 8}$ and ${\bf 10}$ $SU(3)$
multiplets of baryons, followed by an exotic ${\bf \overline {10}}$
multiplet\cite{Manohar,Chemtob,Praszalowicz} whose $S=+1$ member is
the $\Theta^+$.  The fact that the exotic ${\bf \overline {10}}$
multiplet is found simply by exciting the soliton to the next
rotational energy level after the well-known decuplet has led to a
widespread belief that the pentaquarks are a robust prediction of
chiral soliton models, independent of assumptions about the dynamics.
In this talk we argue that this belief is not well-founded.  Instead,
we will conclude that in soliton models the existence or non-existence
of the pentaquarks very much depends on the details of the dynamics,
i.e. the structure of the chiral lagrangian. Thus, the soliton models
do not produce any miracles that are not obvious from general
priciples of QCD. Neither in QCD nor in chiral soliton models is there
anything that a priori guarantees the existence of narrow
pentaquarks. Indeed, we will show that with the standard set of the
Skyrme model parameters, a resonance with the quantum numbers of the
$\Theta^+$ does not form.  This fact should be kept in mind as some of
the more recent searches\cite{nopenta} have failed to confirm the
existence of the $\Theta^+$ or other pentaquarks.

Our theoretical discussion follows the basic premise\cite{Witten} that
the semiclassical quantization of chiral solitons corresponds to
the $1/N$ expansion for baryons in QCD generalized to a large
number of colors $N$.
It is therefore important to generalize the discussion of 
exotic collective coordinate states carried out for $N=3$ in \cite{DPP}
to large $N$.
The allowed multiplets must contain states of hypercharge $N/3$, 
{\it i.e.} of strangeness $S=0$.  In the notation where $SU(3)$ multiplets are
labeled by $(p,q)$, the lowest multiplets 
one finds\cite{Manohar,Vadim,Dul,IRK,Cohen,DP} are $(1,n)$ with
$J=\frac{1}{2}$ and $(3,n-1)$ with $J=\frac{3}{2}$.
These are the large $N$ analogues of the octet and the decuplet. 
Exactly the same multiplets appear when we construct baryon states out
of $N$
quarks. The splittings among them are of order $1/N$,
as is usual for soliton rotation excitations.


The large $N$ analogue of the exotic antidecuplet is the
representation $(0, n+2)$ with $J=\frac{1}{2}$, and one finds that its
splitting from the lowest multiplets is $O(N^0)$ in the rotator
approximation.  The fact that the mass splitting is of order one,
comparable to the energy of mesonic fluctuations, raises
questions\cite{KK,IRK,Cohen} about the validity of the rigid rotator
approach to these states. Instead, a better treatment of these states
is provided by the bound state approach\cite{CK} where one departs
from the rigid rotator ansatz and adopts more general kaon
fluctuation profiles; in this approach one describes the $\Theta^+$ as
a kaon-skyrmion resonance or bound state of $S=+1$, rather than by a
rotator state (a similar suggestion was made independently by
Cohen\cite{Cohen}.)
In the non-exotic sector, as we take the limit $m_K\to 0$ the bound
state description of low-lying baryons smoothly approaches the rigid
rotator description, and the bound state wavefunction approaches a
zero-mode.  However, in contrast with the situation for $S=-1$, for
$S=+1$ there is no fluctuation mode that in the $m_K\to 0$ limit
approaches the rigid rotator mode. Thus, for large $N$ and small 
$SU(3)$ breaking,
the rigid rotator state with $S=+1$ is an artifact of the rigid
rotator approximation (we believe this to be a general statement that
does not depend on the details of the chiral lagrangian.)

Next we ask what happens as we increase the $SU(3)$ breaking by
varying parameters in the effective lagrangian (such as the kaon mass
and the weight of the Wess-Zumino term) and find that a substantial
departure from the $SU(3)$-symmetric limit is necessary to stabilize
the kaon-skyrmion system.  We reach a conclusion that, at least for
large $N$, the exotic $S=+1$ state exists only due to the $SU(3)$
breaking and disappears when the breaking is too weak.

\section{The rigid rotator vs. the bound state approach}

Our discussion of chiral solitons is mainly carried out in the context of
the Skyrme model, but our conclusions will not be tied to a specific model.
The Skyrme approach to baryons begins with the Lagrangian\cite{Skyrme}
\beq
L_{Skyrme} = \frac{f^2_{\pi}}{16} {\rm Tr}(\partial_{\mu} U^{\dag}
\partial^{\mu} U)
&+& \frac{1}{32e^2}{\rm Tr}([\partial_{\mu} U U^{\dag},\partial_{\nu} U
U^{\dag}]^2) \nonumber\\
&+& {\rm Tr}( M (U+U^{\dag}-2)) \, ,
\label{lskyrme}
\eeq
where $U(x^{\mu})$ is a matrix in $SU(3)$ and $M$ is proportional to
the matrix of quark masses. 
There is an additional term in the action,
called the Wess-Zumino term, whose
normalization \cite{Witten} is proportional to $N$.

Skyrme showed that there are topologically
stabilized static solutions of hedgehog form, 
$U_0= e^{i{ {\bf \tau} \cdot \hat{r}}F(r)}$,
%
%
in which the radial profile function $F(r)$ satisfies the boundary
conditions $F(0)=\pi, F(\infty)=0$.
The non-strange low-lying excitations of this soliton are given by
rigid rotations of the pion field $A(t) \in SU(2)$:
\beq
U(x,t)=A(t)U_0 A^{-1}(t).
\label{rot}
\eeq
For this ansatz the Wess-Zumino term does not contribute.
By expanding the Lagrangian about $U_0$ and canonically quantizing the
rotations, one finds that the Hamiltonian is
\beq
H=M_{cl} + \frac{1}{2\Omega} J(J+1) \, ,
\eeq
where $J$ is the spin and the c-numbers $M_{cl}$ and $\Omega$ are
functionals of the soliton profile.
For vanishing pion mass, one finds numerically that
\beq
M_{cl} \simeq& 36.5 \frac{f_{\pi}}{e}\ ,\qquad
\Omega \simeq& \frac{107}{e^3 f_{\pi}} \, .\label{omega}
\eeq
For $N=2n+1$, the low-lying quantum numbers are independent of the
integer $n$.  The lowest states, with $I=J=\frac{1}{2}$ and
$I=J=\frac{3}{2}$, are
identified with the nucleon and $\Delta$ particles respectively. Since
$f_\pi\sim \sqrt{N}$, and $e\sim 1/\sqrt{N}$, the soliton mass is
$\sim N$, while the rotational splittings are $\sim 1/N$.  Adkins,
Nappi and Witten\cite{Adkins} found that they could fit the $N$ and
$\Delta$ masses with the parameter values $e=5.45, f_{\pi}=129$ MeV.
In comparison, the physical value of $f_{\pi}=186$ MeV.

A generalization of this rigid rotator treatment that produces $SU(3)$
multiplets of baryons is obtained by making the collective coordinate
$A(t)$ an element of $SU(3)$.  Then the WZ term makes a crucial
constraint on allowed multiplets\cite{Witten,Guad,Manohar,Chemtob}.
The large-$N$ treatment of this 3-flavor
Skyrme model is
more subtle than in the 2-flavor case. When $N=2n+1$ is large,
even the lowest lying $(1,n)$ $SU(3)$ multiplet contains $(n+1)(n+3)$
states with strangeness ranging from $S=0$ to $S=-n-1$
\cite{IRK}. When the strange quark mass is turned on, it
introduces a splitting of order $N$ between the lowest and highest
strangeness baryons in the same multiplet.  Thus, $SU(3)$ is badly
broken in the large $N$ limit, no matter how small $m_s$ is
\cite{IRK}.  It is helpful to think in terms of $SU(2)\times
U(1)$ flavor quantum numbers, which do have a smooth large $N$ limit.
In other words, we focus on low strangeness members of these
multiplets, whose $I$, $J$ quantum numbers have a smooth large $N$
limit, and identify them with observable baryons.

Since the multiplets contain baryons with up to $\sim N$ strange
quarks, the wave functions of baryon with fixed strangeness deviate
only an amount $\sim 1/N$ into the strange directions of the
collective coordinate space. Thus, to describe them, one may expand
the $SU(3)$ rigid rotator treatment around the $SU(2)$ collective
coordinate.  The small deviations from $SU(2)$ may be assembled into a
complex $SU(2)$ doublet $K(t)$.  This method of $1/N$ expansion was
implemented in \cite{KK}, and reviewed in \cite{IRK}.


{}From the point of view of the Skyrme model the ability to expand in
small fluctuations is due to the Wess-Zumino term which acts as a large
magnetic field of order $N$. The method works for arbitrary kaon mass,
and has the correct limit as $m_K\to 0$.  To order $O(N^0)$ the
Lagrangian has the form\cite{KK}
\begin{equation}
L = 4\Phi \dot K^\dagger \dot K + i {N\over 2} (K^\dagger \dot K - \dot
K^\dagger  K )
- \Gamma K^\dagger K \, .
\end{equation}
The Hamiltonian may be diagonalized:
\begin{equation}
H= \omega_- a^\dagger a +
\omega_+ b^\dagger b + {N\over 4\Phi}
\ ,
\end{equation}
where
\begin{equation}
\omega_\pm = {N\over 8\Phi} \left (\sqrt{1+ (m_K/M_0)^2} \pm 1 \right )
\ ,\qquad M_0^2 = {N^2\over 16 \Phi\Gamma }\ .
\end{equation}
The strangeness operator is $S= b^\dagger b - a^\dagger a$.  All the
non-exotic multiplets contain $a^\dagger$ excitations only. In the
$SU(3)$ limit, $\omega_-\to 0$, but $\omega_+ \to {N\over 4\Phi}\sim
N^0$.  Thus, the ``exoticness'' quantum number mentioned in \cite{DP}
is simply $E= b^\dagger b$ here, and the splitting between multiplets
of different ``exoticness'' is ${N\over 4\Phi}$, in agreement with
results found from the rigid rotator\cite{Vadim,Dul,IRK,Cohen,DP}.


The $O(N^0)$ splittings predicted by the rigid rotator are, 
however, not exact: this approach does not
take into account deformations of the soliton 
as it rotates in the strange directions\cite{CK,KK,IRK}.  
Another approach to strange baryons, which allows for these deformations, 
and which proves to be quite
successful in describing the light hyperons, is the so-called bound
state method\cite{CK}.  The basic strategy is to expand the
action to second order in kaon fluctuations about the classical
hedgehog soliton.  Then one can obtain a linear differential equation
for the kaon field, incorporating the effect of the kaon mass, which
one can solve exactly.  The eigenenergies of the kaon field are then
the $O(N^0)$ differences between the masses of the strange
baryons and the classical Skyrmion mass. 
It is convenient to
write $U$ in the form
$ U=\sqrt{U_{\pi}} U_K \sqrt{U_{\pi}}$,
where $U_{\pi}=\exp[2i\lambda_j \pi^j/f_{\pi}]$ and
$U_K=\exp[2i\lambda_a K^a/f_{\pi}]$ with $j$ running from 1 to 3 and
$a$ running from 4 to 7. The $\lambda_a$ are the
standard $SU(3)$ Gell-Mann matrices.  We will collect the $K^a$ into a
complex isodoublet $K$:
\beq
K=
\frac{1}{\sqrt{2}}\left(
\begin{array}{c}
K^4-iK^5 \\
K^6-iK^7
\end{array} \right)=
\left(
\begin{array}{c}
K^+ \\K^0
\end{array} \right).
\label{kdoublet}
\eeq
Expanding the Wess-Zumino term to second order in $K$, we obtain 
\beq
L_{WZ} = \frac{iN}{f_{\pi}^2}B^{\mu} \left( K^{\dag}
D_{\mu}K-(D_{\mu}K)
^{\dag}K \right)
\eeq
where
\beq
D_{\mu}K=
\partial_{\mu}K+\frac12\left(\sqrt{U_{\pi}^{\dag}}
\partial_{\mu}\sqrt{U_{\pi}}+\sqrt{U_{\pi}}
\partial_{\mu}\sqrt{U_{\pi}^{\dag}}\right)K \, ,
\eeq
and $B_\mu$ is the baryon number current. Now we decompose
the kaon field into a set of partial waves.  Because
the background soliton field is invariant under combined spatial and
isospin rotations ${\bf T} = {\bf I}+{\bf L}$, a good set of quantum
numbers is $T,L$ and $T_z$, and so we write the kaon eigenmodes as
$ K=k(r,t) Y_{TLT_z} $.
Substituting this expression into $L_{Skyrme}+L_{WZ}$ we obtain an
effective Lagrangian for the radial kaon field $k(r,t)$ of the form\cite{CHK}
\beq \nonumber
L=4\pi \int r^2 dr\Bigg(f(r)\dot{k}^{\dag}\dot{k}+ i\lambda(r)(k^{\dag}\dot{k}&-&\dot{k}^{\dag}k)\nonumber \\
-h(r)\frac{d}{dr}k^{\dag}\frac{d}{dr}k &-&k^{\dag}k(m_K^2+V_{\rm eff}(r)) \Bigg) \, .
\eeq
The formula for the effective potential $V_{\rm eff}(r)$ appears in \cite{CHK,us}.
The resulting equation of motion for $k$ is
\beq
-f(r)\ddot{k} +2i\lambda(r)\dot{k}+\mathcal{O} k=0\, , \\
\mathcal{O} \equiv \frac{1}{r^2}\partial_r
h(r)r^2\partial_r-m_K^2-V(r)\, . \nonumber
\eeq
The eigenvalue equations are
\beq
(f(r)\omega_n^2+2\lambda(r)\omega_n+\mathcal{O} )k_n&=&{ 0} \qquad(S=-1)\,
,\nonumber \\
(f(r)\tilde{\omega}_n^2-2\lambda(r)\tilde{\omega}_n+\mathcal{O})\tilde{k}_n&=&
{ 0}
\qquad (S=+1)\, ,
\label{keqns}
\eeq
with $\omega_n,\tilde{\omega}_n$ positive.
Crucially, the sign in front of $\lambda$, which is the contribution
of the WZ term, depends on whether the relevant eigenmodes have
positive or negative strangeness.  


It is possible to examine these equations analytically for $m_K=0$.
Then one finds that the $S=-1$ equation has an exact solution with
$\omega=0$ and $k(r)\sim \sin (F(r)/2)$, which corresponds to the
rigid rotator zero-mode\cite{CHK}. As $m_K$ is turned on, this
solution turns into an actual bound state \cite{CK,CHK}. On the other
hand, the $S=+1$ equation does {\it not} have a solution with $\tilde
{\omega} ={N\over 4\Phi}$ and $k(r)\sim \sin (F(r)/2)$. This is why
the exotic rigid rotator state is not reproduced by the more precise
bound state approach to strangeness. In section 3 we further check
that, for small $m_K$, there is no resonance corresponding to the
rotator state of energy ${N\over 4\Phi}$ in the $SU(3)$ limit.


The lightest $S=-1$ bound state is in the channel $L$=1,
$T=\frac{1}{2}$, and its mass is $M_{cl}+0.218\, ef_{\pi} \simeq 1019$
MeV.  This state gives rise to the $\Lambda(1115)$, $\Sigma(1190)$,
and $\Sigma(1385)$ states, where the additional splitting arises from
$SU(2)$ rotator corrections\cite{CK,CHK}.  There is also a $L=0$,
$T=\frac{1}{2}$ bound state corresponding to the negative parity
hyperon $\Lambda(1405)$.  The natural appearance of the
$\Lambda(1405)$ is a major success of the bound state
approach\cite{CHK,Gobbi}.

\begin{figure}[htbp]
\centering
\epsfig{file=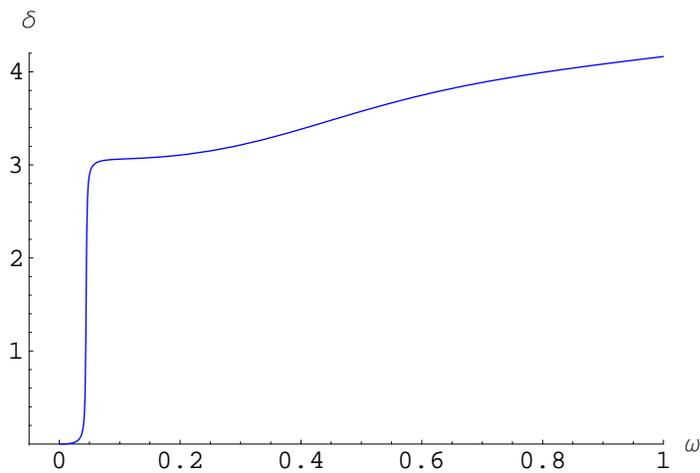, width=4in}
\caption{Phase shift as a function of energy in the
$L=2$, $T=\frac{3}{2}$, $S=-1$ channel.
The energy $\omega$ is measured in units of $ef_\pi$ (with the kaon mass
subtracted, so that $\omega=0$ at threshold), and
the phase shift $\delta$ is measured in radians.  Here $e=5.45$ and
$f_\pi=129$ MeV.}
\label{dwavefig}
\end{figure}


The same method can be applied also to states above threshold.  Such
states will appear as resonances in kaon-nucleon scattering, which we
may identify by the standard procedure of solving the appropriate kaon
wave equation and studying the phase shifts of the corresponding
solutions as a function of the kaon energy. In the $L=2$,
$T=\frac{3}{2}$ channel there is a resonance at $M_{cl}+0.7484 \,
ef_{\pi}$=1392 MeV (see Figure 1).
Upon the $SU(2)$ collective coordinate quantization, it gives rise to
three states\cite{Scoc} with $(I, J)$
given by  $(0, \frac{3}{2})$, $(1, \frac{3}{2})$,
$(1, \frac{5}{2})$,
 with masses 1462 MeV, 1613
MeV, and 1723 MeV respectively (see Table 2).
We see that these correspond nicely to the known negative
parity resonances $\Lambda(1520)$ (which is $D_{03}$ in standard
notation),
$\Sigma(1670)$ (which is $D_{13}$) and
$\Sigma( 1775)$ (which is $D_{15}$).
As with the bound states, we find that the resonances are
somewhat overbound (the overbinding of all states is presumably
related to the necessity of adding an overall zero-point energy
of kaon fluctuations), but
that the mass splittings within this multiplet are accurate to within a
few percent. In fact, we find that the ratio
\beq
{M(1, \frac{5}{2})- M(0, \frac{3}{2})\over
M(1, \frac{3}{2}) - M(0, \frac{3}{2})}
\approx 1.73
\ ,
\eeq
while its empirical value is $1.70$.

\section{Baryons with S=+1?}

For states with positive strangeness, the eigenvalue equation for the
kaon field is the same except for a change of sign in the contribution
of the WZ term.  This sign change makes the WZ term repulsive for
states with $\bar{s}$ quarks and introduces a splitting between
ordinary and exotic baryons\cite {CK}.  In fact, with standard values
of the parameters (such as those in the previous section) the
repulsion is strong enough to remove {\em all} bound states and
resonances with $S=1$, including the newly-observed
$\Theta^+$.  It is natural to ask how much we must modify the Skyrme
model to accommodate the pentaquark.  The simplest modification we can
make is to introduce a coefficient $a$ multiplying the WZ term.
Qualitatively, we expect that reducing the WZ term will make the
$S=+1$ baryons more bound, while the opposite should happen to the
ordinary baryons.  


The most likely channel in which we might find an exotic has the
quantum numbers $L=1$, $T=\frac{1}{2}$, as in this case the effective
potential is least repulsive near the origin.  For $f_{\pi}$= 129,
186, and 225 MeV, with $e^3f_{\pi}$ fixed, we have studied the effect
of lowering the WZ term by hand.  Interestingly, in all three cases we
have to set $a\simeq 0.39$ to have a bound state at threshold.  If we
raise $a$ slightly, this bound state moves above the threshold, but
does not survive far above threshold; it ceases to be a sharp state
for $a\simeq 0.46$.  We have plotted phase shifts for various values
of $a$ in Figure 2.\footnote{When the state is above the threshold, we
do not find a full $\pi$ variation of the phase. Furthermore, the
variation and slope of the phase shift decrease rapidly as the state
moves higher, so it gets too broad to be identifiable. So, the state
can only exist as a bound state or a near-threshold state.  }
Assuming that the parameters of the chiral lagrangian take values such that the $\Theta^+$ exists, we can then consider the $SU(2)$
collective coordinate
quantization of the state, in a manner analogous to the treatment of the $S=-1$
bound states\cite{CHK}.  Here we record our results, assuming that $a=0.39$ and $f_\pi=129$ MeV, and refer the reader to our original paper\cite{us} for details.

The lightest $S=+1$ state we find has $I = 0$, $J = \frac{1}{2}$ and
positive parity, {\it i.e.} it is an $S= +1$ counterpart of the
$\Lambda$.  This is the candidate $\Theta^+$ state.  Its first $SU(2)$
rotator excitations have $I =1$, $J^P = \frac{3}{2}^+$ and $I =1$,
$J^P = \frac{1}{2}^+$ (a relation of these states to $\Theta^+$ also
follows from general large $N$ relations among baryons\cite{Cohen,Jen}).
The counterparts of these $J^P=\frac{1}{2}^+, \frac{3}{2}^+$ states in
the rigid rotator quantization lie in the {\bf 27}-plets of $SU(3)$
\cite{Walliser,Kobushkin}.
We find that the $I=1$, $J^P = \frac{3}{2}^+$ state is $\sim 148$ MeV
heavier than the $\Theta^+$,
while the $I =1$, $J^P = \frac{1}{2}^+$
state is $\sim 289$ MeV
heavier than the $\Theta^+$.

We may further consider $I=2$ rotator excitations which have
$J^P= \frac{3}{2}^+, \frac{5}{2}^+$.
Such states are allowed for $N=3$ (in the quark
language the charge $+3$ state, for example, is given by
$uuuu{\bar s}$).
The counterparts of these
$J^P=\frac{3}{2}^+, \frac{5}{2}^+$
states in the rigid rotator quantization lie in the {\bf 35}-plets of
$SU(3)$\cite{Walliser,Kobushkin}. We find
\beq
& M(2, \frac{5}{2})- M(0, \frac{1}{2}) \sim 494
\ {\rm MeV}, \nonumber \\
& M(2, \frac{3}{2})- M(0, \frac{1}{2}) \sim 729\ {\rm MeV}\ .
\eeq

Although the specific mass splittings which we have computed depend on
the choice of parameters in the chiral lagrangian, it turns out that
we may form certain combinations of masses of the exotics which rely
only the existence of the $SU(2)$ collective coordinate:
\beq \label{modelin}
& 2 M(1, \frac{3}{2}) + M(1, \frac{1}{2})- 3 M(0, \frac{1}{2})
 =
2(M_\Delta- M_N)=586\ {\rm MeV} \ \nonumber ,\\
& \frac{3}{2} M(2, \frac{5}{2}) + M(2, \frac{3}{2})- \frac{5}{2}
M(0, \frac{1}{2}) = 5(M_\Delta- M_N)= 1465\ {\rm MeV}\ , \nonumber \\
&   M(2,\frac{3}{2}) - M(2,\frac{5}{2}) = \frac{5}{3} \left(
M(1,\frac{1}{2})-  M(1,\frac{3}{2}) \right) , 
\eeq
where we used $M_\Delta-M_N=\frac{3}{2\Omega}$. These
``model-independent'' relations have also been derived using a different method
\cite{Jen}.

\begin{figure}[htbp]
\centering
\epsfig{file=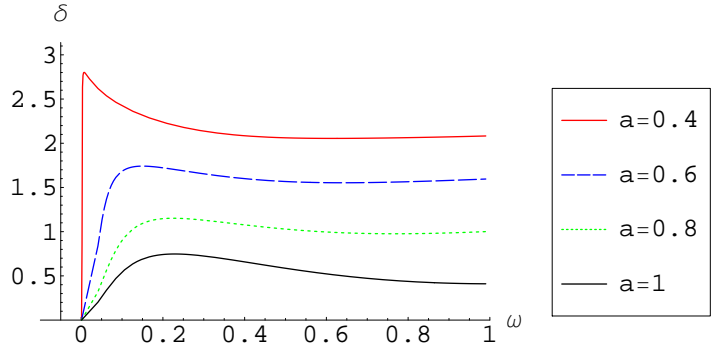, width=4in}
\caption{Phase shifts $\delta$ as a function of energy
in the $S=+1$, $L=1$, $T=\frac{1}{2}$ channel, for
various choices of the parameter $a$ (strength of the WZ term).
 The energy $\omega$ is measured in units of $ef_\pi$ ($e=5.45$,
$f_\pi=f_K= 129$ MeV) and
the phase shift $\delta$ is measured in radians. $\omega =0$
corresponds to the $K-N$ threshold.}
\label{wzplot}
\end{figure}


As another probe of the parameter space of our Skyrme model, we may
vary the mass of the kaon and see how this affects the pentaquark.  As
observed in Section 3, in the limit of infinitesimal kaon mass, there
is no resonance in the $S=+1, L=1, T=\frac{1}{2}$ channel.  We find
that to obtain a bound state in this channel, we must raise $m_K$ to
about 1100 MeV. Plots of the phase
shift vs. energy for different values of $m_K$ may be found in \cite{us}.

\section{Discussion}

The main implication of our analysis is that in chiral soliton
models there is no ``theorem'' that exotic pentaquark baryons exist,
nor is there a theorem that they do not exist. The situation really depends
on the details of the dynamics inherited from the underlying QCD.

The statements above apply to general chiral soliton models containing various
lagrangian terms consistent with the symmetries of low-energy hadronic physics.
In the bound state approach to the Skyrme model we saw that
an $S=+1$ near-threshold state is absent when we use the standard parameters,
but comes into existence only at the expense of a large reduction in the Wess-Zumino 
term. It is doubtful that such a reduction is consistent with QCD.
However, one can and should explore other variants of chiral soliton models.
For example, in \cite{Rhonew} the exotic $S=+1$ resonances were studied
in a model containing explicit $K^*$ fields. This model contains a coupling constant
which is, roughly speaking, the analogue of the coefficient of the WZ term, $a$, in 
our approach. The findings of \cite{Rhonew} are largely parallel to ours.
For a wide range of values of this coupling, the repulsion is too strong,
and no $S=+1$ resonances can form.
When this coupling is very small, then there exists an $S=+1$ bound state.
There is also a narrow intermediate range where this bound state turns into a 
near-threshold resonance. 
An important question is whether chosing parameters to lie in this narrow  range
is consistent with the empirical constrains on the effective lagrangian.
If not, then one may have to conclude that chiral soliton models actually
predict the {\it absence} of pentaquarks.

\section*{Acknowledgments}
We are grateful to N. Itzhaki and L. Rastelli for collaboration
on a paper reviewed here, and to E. Witten for discussions.
This material is based upon work
supported by the National Science Foundation Grants No.
PHY-0243680 and PHY-0140311.
Any opinions, findings, and conclusions or recommendations expressed in
this material are those of the authors and do not necessarily reflect
the views of the National Science Foundation.

\end{document}